\documentclass[11pt]{JHEP3}

\usepackage{graphicx}
\usepackage{amsmath}
\usepackage{amssymb}

\newcommand{\be}{\begin{eqnarray}}
\newcommand{\ee}{\end{eqnarray}}
\newcommand{\nn}{\nonumber}
\newcommand{\bn}{\begin{enumerate}}
\newcommand{\en}{\end{enumerate}}

\def\IC{\mathbb{C}}

\def\IR{\mathbb{R}}
\def\IZ{\mathbb{Z}}

\def\CN{{\cal N}}

\def\a{\alpha}
\def\b{\beta}

\def\d{\delta}

\def\th{\theta}

\def\D{\Delta}

\def\S{\Sigma}

\def\half{\frac{1}{2}}

\def\goto{\rightarrow}

\def\det{{\rm det}}

\title{Toric AdS$_\mathbf{4}$/CFT$_\mathbf{3}$ duals and M-theory Crystals}

\author{Sangmin Lee$^1$, Sungjay Lee$^1$ and Jaemo Park$^2$\\
\\
$^1$School of Physics and Astronomy, Seoul National University, 
Seoul 151-747, Korea
{}\\
$^2$Department of Physics, Postech, 
Pohang 790-784, Korea
}
\abstract{
We study the recently proposed crystal model for three dimensional 
superconformal field theories arising from M2-branes probing 
toric Calabi-Yau four-fold singularities. 
We explain the algorithms mapping a toric Calabi-Yau to a crystal
and vice versa, and show how the spectrum of BPS meson states 
fits into the crystal model.}


\preprint{hep-th/0702120\\
SNUST-070201}

\begin{document}

\section{Introduction}

The brane tiling model \cite{t1, t2, t3, t4, t5, t6}, 
a.k.a. dimer model, has proved to 
be a powerful tool to analyze the class of four dimensional, 
superconformal quiver gauge theories 
arising from D3-branes probing toric Calabi-Yau three-fold (CY$_3$) cones. 
Over the past few years, systematic methods that map a brane tiling to 
the associated toric CY$_3$ cone (forward algorithm \cite{t1}) and vice versa 
(inverse algorithm \cite{t4,t5}) have been firmly established. 
The brane tiling model has been used to study aspects of 
the quiver gauge theories such as marginal deformations 
\cite{butti, zaff, ima2}, counting BPS states 
\cite{msy2, pleth1, ms2, man2, baryon1, romel, grant, baryon2, pleth2}, 
and partial Higgsing \cite{t2, mp, jaemo, uranga}. 
It has also found interesting applications to the
recent study of supersymmetry breaking of metastable vacua
and their embedding into string theory \cite{uranga2,uranga3,uranga4}.

\begin{figure}[htb]
\begin{center}
\includegraphics[width=8.0cm]{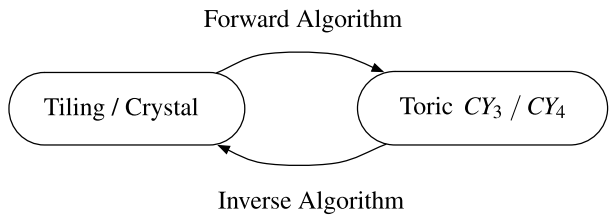}
\caption{The forward/inverse algorithms.} \label{intro}
\end{center}
\end{figure}

Recently, the idea of the brane tiling model was applied 
to three dimensional superconformal theories arising from 
M2-banes probing toric CY$_4$ cones \cite{crystal}. 
A three dimensional counterpart of the tiling model emerged and 
was named a crystal model. 
Throughout this paper, we will distinguish the D3/CY$_3$ model 
from the M2/CY$_4$ model by referring to the former as the tiling model 
and the latter as the crystal model. The notion of a dimer is equally relevant 
to both models. 

The goal of this paper is to elaborate on the properties of the crystal model. 
We begin with a brief review of the construction of the crystal model 
\cite{crystal} in section 2. 
In sections 3 and 4, we generalize the inverse/forward algorithms 
of the tiling model to the crystal model.  
We find that the forward algorithm can be carried over to the crystal model 
with little modification. 
The inverse algorithm also shares many essential features with 
that of the tiling model, 
but there are some crucial differences that require additional work. 
In section 5, we show in detail how the crystal model correctly reproduces 
the spectrum of the BPS meson states.
We conclude with some discussions in section 6.

\section{Review of the Crystal Model}

We begin with a stack of $N$ M2-branes near the tip of a CY$_4$ cone $X$. 
In the near horizon limit, the $\IR^{1,2}$ world-volume directions 
of the M2-branes 
and the radial direction of $X$ merge to form an AdS$_4$ 
and the base (unit radius section) of $X$ becomes 
the internal 7-manifold $Y$. We say $X$ is the cone over $Y$, or $X=C(Y)$. 
%The cone $X$ being K\"ahler and Ricci-flat 
%is equivalent to the base $Y$ being Sasakian and Einstein, respectively. 
The crystal model assumes that $X$ is toric, so we first recall 
some relevant aspects of toric geometry. 
See, for example, \cite{ms1, msy1, baryon1} for more information 
on toric geometry in this context. 

\subsection{Toric CY$_\mathbf{4}$}

We follow the notation of \cite{crystal, LR} for toric geometry. 
The CY$_4$ cone X is a quotient of $\IC^d$ for some $d\ge 4$.
Given some integer-valued charge matrix 
$Q_a^I$ $(I=1, \cdots, d; a=1, \cdots, d-4)$, the quotient 
is taken by 
\be 
X = \left\{ \sum_I Q_a^I |Z_I|^2 = 0 \right\} / (Z_I \sim e^{Q_a^I\th^a} Z_I).
\ee
The toric diagram is a convex polyhedron composed of a set of lattice points 
$\{v_I^i\} \in \IZ^4$ ($i=1, \cdots, 4$) satisfying 
\be 
\sum_I Q_a^I v_I^i = 0.  \label{ker}
\ee
The CY condition is $\sum_I Q_a^I=0$, which leads 
the $v_I$ to be on the same $\IZ^3$ subspace. 
It is customary to choose a basis to set $v_I^4=1$ for all $I$. 

The toric diagram defines a solid cone 
$\D \equiv \{ y_i \in \IR^4 ; (v_I \cdot y) \ge 0 \mbox{ for all }I\}$. 
We call the boundary components 
$S^I \equiv \D \cap \{v_I\cdot y = 0\}$ the 3-fans. 
Two 3-fans meet at a 2-fan and several 2-fans join at a 1-fan. 
These fans are graph dual to the original toric diagram in 
the sense that each vertex $v_I$ is associated to a 3-fan, 
each edge connecting two neighboring vertices, $w_{IJ} = v_I -v_J$, 
corresponds to a 2-fan, etc. 
The cone $X$ is a $T^4$ fibration over $\D$. 
The fiber is aligned with the base in such a way that it shrinks to $T^n$ on 
the $n$-fans.

\begin{figure}[htb]
\begin{center}
\includegraphics[width=8.5cm]{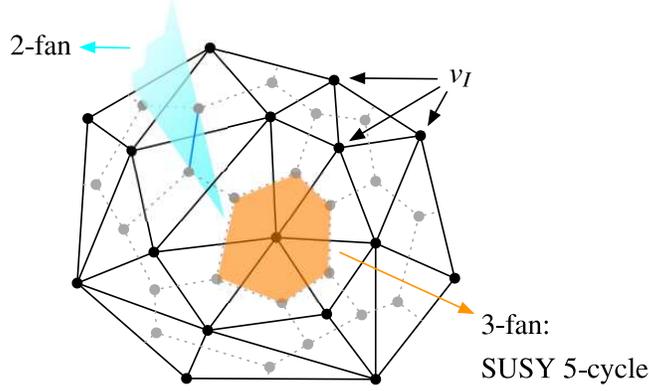}
\caption{A toric diagram (solid line) is
a convex polyhedron with integer-valued vertices in $\IR^3$. 
Its graph dual (dotted line) gives the fan diagram.} \label{partition}
\end{center}
\end{figure}

The moduli space of K\"ahler metrics on $X$ is parameterized by the Reeb vector 
$b^i \in \IR^4$, which defines the base of the cone by 
$Y = X \cap\{b\cdot y = 1/2\}$. 
In the basis mentioned above, the CY condition fixes $b^4=4$. 
The volume of $Y$ as an explicit function of $v_I$ and $b$ 
is known \cite{msy1, msy2}. 
The Ricci-flat metric is obtained by minimizing the volume with respect to 
$b^i$ with the range of $b^i/4$ being precisely the toric diagram. 

\begin{figure}[htb]
\begin{center}
\includegraphics[height=17.0cm]{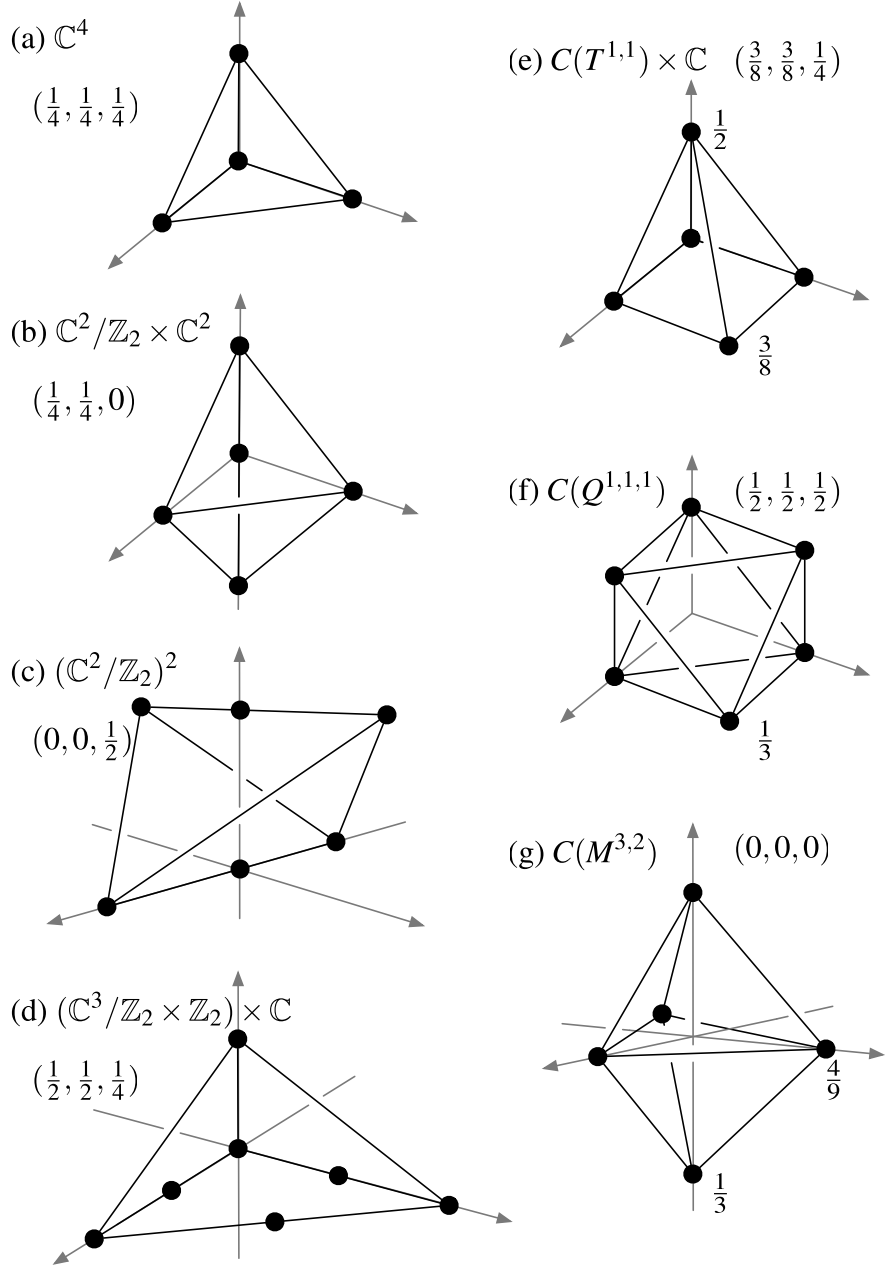}
\caption{Toric diagrams, Reeb vectors and R-charges of 
some simple examples of CY$_4$.} \label{xamp}
\end{center}
\end{figure}

For later purposes, we recall that a baryon in the CFT$_3$ is mapped 
via AdS/CFT to an M5-brane wrapping a supersymmetric 5-cycle of $Y$ 
\cite{bhk}.
Each 3-fan $S^I$ dual to a vertex $v_I$ defines a 5-cycle 
which is the $T^3$ fibration over $S^I \cap \{b\cdot y = 1/2\}$. 
We abuse the notation a bit and use $S^I$ to denote either 
the 3-fan or the 5-cycle depending on the context. 
%The baryons are charged under the global symmetries of the theory. 
%The four $U(1)$ isometries $F_i$ of $X$ are called flavor symmetries. 
It was shown in \cite{t3} that $H_5(Y,\IZ)=\IZ^{d-4}$ with $d$
  being the number of internal vertices in the toric diagram, and that 
a set of basis $\{{\cal C}^a\} \subset H_5(Y,\IZ)$ can be chosen such that that 
$S^I=Q_a^{I} {\cal C}^a$, where $Q_{a}^{I}$ is precisely the charge matrix 
defining the CY$_4$. The baryons $S^I$ are said to be charged under the baryonic symmetries $Q_a$. The baryons are also charged
under the four $U(1)$ isometries $F_i$ of $X$, called the flavor
symmetries.
There are also $d-4$ baryonic symmetries $Q_a$.
%where $d$ is the 
%number of vertices of the toric diagram. 
We normalize the charges of the the baryons to be
$F_i[S^I]  \equiv N F_i^I$, $Q_a[S^I] \equiv N Q_a^I$ with $N$ being 
the number of M2-branes.
They satisfy the toric relations $v_I^i F^I_j = \d^i_j$ and $v_I^i Q_a^I = 0$ \cite{LR}.
The $U(1)_R$ charge is the linear combination 
of the flavor charges $R = \half b^i F_i$, where $b^i$ is the Reeb vector. 
The toric relations and the CY condition ($b^4=4$) implies that $\sum_I R^I = 2$ 
and $\sum_I F_i^I = 0$ for $i=1,2,3$. 
The value of each $R^I$ is proportional to the volume of $S^I$ \cite{bhk} 
and can be computed following \cite{msy1, baryon1}.

Figure \ref{xamp} lists the toric diagrams for 
the examples of CY$_4$ we will consider in this paper. 
The orbifolds in examples (b,c,d) are defined as follows.  
For $\IC^2/\IZ_2$, the $\IZ_2$ group acts as $(z_1,z_2)\goto (-z_1,-z_2)$. 
For $(\IC^3/\IZ_2\times \IZ_2)\times \IC$, 
the two $\IZ_2$ groups act on $\IC^3$ as 
$(z_1, z_2, z_3) \goto (-z_1,-z_2,z_3)$ and $(z_1, z_2, z_3) \goto (z_1,-z_2,-z_3)$. 
In example (d), $C(T^{1,1})$ is the famous conifold whose base $T^{1,1}$ 
is the coset space $SU(2)\times SU(2) / U(1)$. 
Examples (e) and (f) are CY$_4$ counterparts of the conifold, 
in the sense that the base $Q^{1,1,1}$ and $M^{3,2}$ 
are coset spaces $SU(2)^3/U(1)^2$ and $SU(3)\times SU(2)/ SU(2)\times U(1)$ 
\cite{cas1}. 
They have been studied in the context of AdS/CFT correspondence 
in \cite{cas2, ak, dall, fab, ahn1, gmsw2, ahn2, glmw}.

In Figure \ref{xamp}, we included the Reeb vector components 
$b^i/4$ ($i=1,2,3$) and the $R$-charges $R^I$ of the baryons $S^I$. 
%The discrete symmetries of the examples simplify the computation enormously. 
For $\IC^4$ and its orbifolds, $R^I=1/2$ for the four corners the toric diagrams 
and zero for other vertices. The $R^I$ for the other three examples 
are written in the diagrams. Note that the vertices related by discrete symmetries have the same $R^I$.

\subsection{The crystal model}

The crystal model follows from a T-duality of M-theory. 
It parallels the derivation of the tiling model 
using T-duality of IIB string theory \cite{t2, ima1}.
We take the T-duality transformation along $T^3\subset T^4$ aligned with the
$y_{1,2,3}$ coordinates. This corresponds to $x^{6,7,8}$ directions in Table 1 below.
By T-duality, we mean the element $t$ in the $SL(2,\IZ)\times SL(3,\IZ)$ duality 
group which acts as $t: \tau \equiv C_{(3)} + i \sqrt{g_{T^3}} \goto -1/\tau$.
The stack of $N$ M2-branes turns into a stack of $N$ M5-branes wrapping 
the dual $T^3$. 
We call them the $T$-branes. 
The degenerating circle fibers turn into another M5-brane 
extended along the (2+1)d world-volume and a non-trivial 3-manifold 
$\S$ in $\IR^3 \times T^3$. We call it the $\S$-brane. 
The result is summarized in Table 1.

\begin{table}[htb]
\label{brane3}
$$
\begin{array}{l|ccc|cccccc|cc}
\hline
           & 0 & 1 & 2 & 3 & 4 & 5 & 6 & 7 & 8 & 9 & 11\\
\hline
\mbox{M5} & \circ & \circ & \circ & & & & \circ & \circ & \circ \\
\mbox{M5} & \circ & \circ & \circ &  \multicolumn{6}{c|}{\Sigma} & & \\
\hline
\end{array}
$$
\caption{The brane configuration for the CFT$_3$.  
Away from the origin of $\IR^3$(345), 
the special Lagrangian manifold $\Sigma$ is locally a product of 
a 2-plane in $\IR^3$(345) and a 1-cycle in $T^3$(678).}
\end{table}

The brane configuration preserves supersymmetry if and only if 
the M5-branes wrap special Lagrangian submanifolds of  
$\IR^3 \times T^3 = (\IC^*)^3$. Clearly, the T-branes are special Lagrangian 
as it is calibrated by $\mbox{Im}\Omega$, where $\Omega$ is the 
holomorphic three-form 
$\Omega = (dx^3+idx^6)\wedge(dx^4+idx^7)\wedge(dx^5+idx^8)$. 
Demanding that the $\S$-brane be calibrated by 
the same $\mbox{Im}\Omega$,  we learn that $\S$ is locally 
a plane in $\IR^3$ and a 1-cycle in $T^3$. 
We will say more about the geometry of $\S$ in section 3. 

In analogy with the tiling model, 
the content of the CFT$_3$ is expected to be encoded 
in the intersection locus between 
the $T$-branes and the $\S$-brane projected onto the $T^3$. 
The result is a graph in the $T^3$ which we call the crystal lattice.  
As in the tiling model, the graph consists of edges and 
vertices which we call {\em bonds} and {\em atoms} 
to distinguish them from similar objects in the toric diagram. 
In \cite{crystal}, the simplest examples of the crystal were presented, 
and a few general properties of the inverse algorithm were stated 
without detailed explanation. We will study both the inverse and 
forward algorithms in sections 3 and 4, respectively, illustrating 
the ideas with the examples listed in Figure \ref{xamp}.

One of the central result of \cite{crystal} is that 
the fundamental excitations of the CFT$_3$ are the 
M2-discs whose boundary encircles the bonds of the crystal lattice.
The derivation of the M2-disc picture made use of the 
smooth transition from a baryon to $N$ fundamentals, 
inspired by a related work in the tiling model \cite{ima1}. 
Using this picture, the super-potential terms 
and the F-term conditions of the crystal model were identified.

As an application of the crystal model, it was shown in \cite{crystal} 
how to account for the spectrum of BPS meson operators using 
the M2-disc picture. In the original CY$_4$, the mesons are KK momentum 
modes. The T-duality transforms them into closed membranes. 
The R-charge and the flavor charges of a meson can be determined 
by decomposing it into elementary M2-discs. 
It is consistent with the known results on the BPS spectrum of chiral 
mesons from the geometry side \cite{msy2, pleth1, ms2}. 
In section 5, we will explain the details of this matching 
with explicit computations for a few examples. 

\section{Inverse algorithm}

In this section, we generalize 
the inverse algorithm of the tiling model \cite{t4, t5} to the crystal model. 
The notion of amoeba and alga projections again plays an important role. 

\subsection{Review of the tiling model}

In the tiling model, 
the T-duality maps the degenerating circle fibers into an NS5-brane 
wrapping a holomorphic surface $\S$ in $(\IC^*)^2$. 
The surface is given by the Newton polynomial, 
\be
\label{newton}
\sum_{(a,b)} c_{(a,b)} u^a v^b = 0, \;\;\;\;\; (u,v)\in (\IC^*)^2.
\ee 
where the sum runs over the vertices of the toric diagram. 
The projection of $\S$ onto $\IR^2$ is called the amoeba and the projection onto $T^2$ is called the alga \cite{t5}. 
The amoeba is a thickened $(p,q)$-web. Each leg of the $(p,q)$-web 
asymptotes to a cylinder. If the legs get pulled in toward the center, 
the amoeba can be regarded as  a punctured Riemann surface. 
The genus of the Riemann surface 
is equal to the number of internal points of the toric diagram \cite{t5}.

\begin{figure}[htb]
\begin{center}
\includegraphics[width=8.0cm]{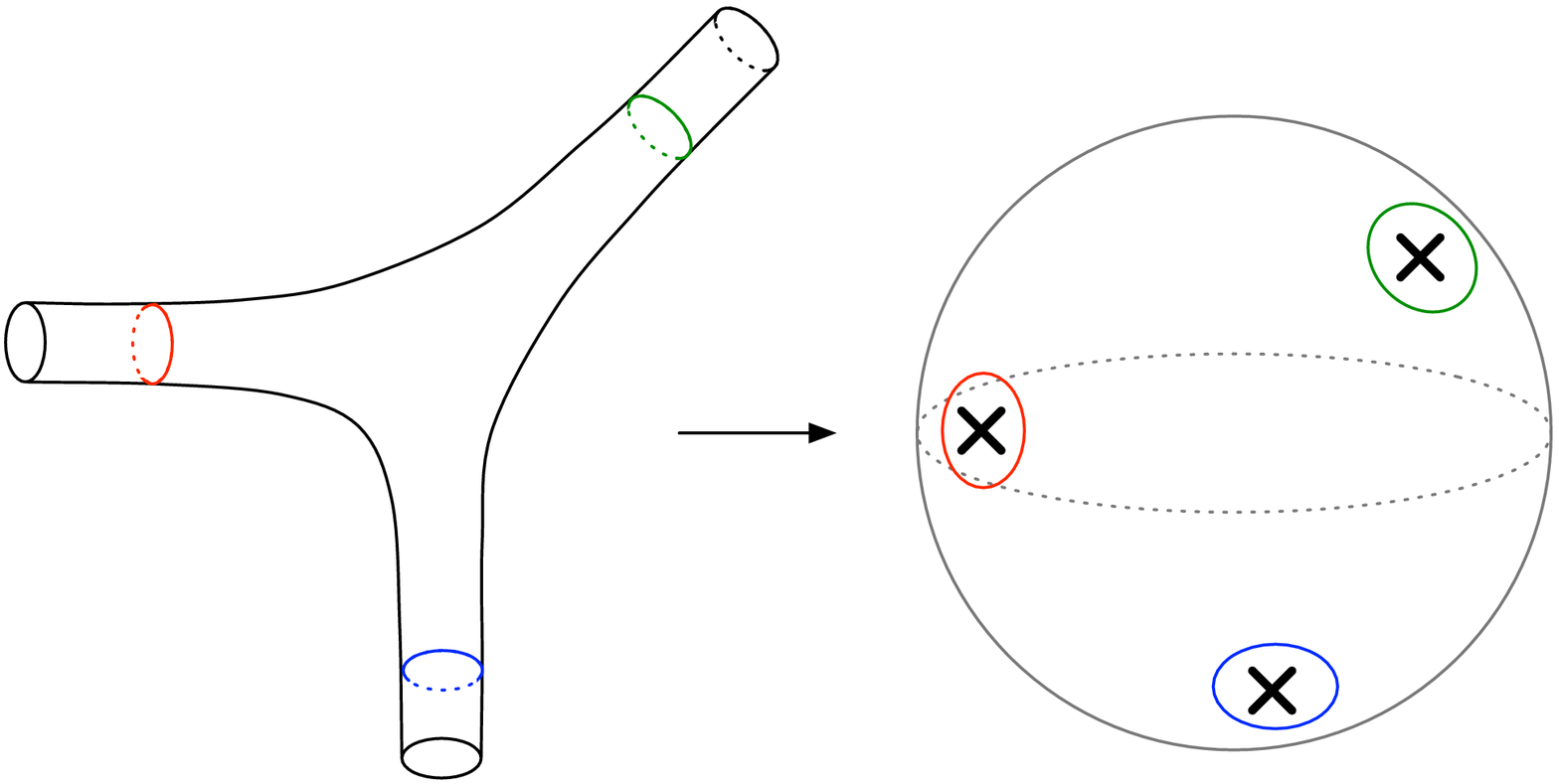}
\caption{Amoeba for $\IC^3$.} \label{amoeba1}
\end{center}
\end{figure}

The alga provides the key to the inverse algorithm. 
The boundaries of the alga are the 1-cycles associated to the 
legs of $(p,q)$-web. The tiling is obtained as the 1-cycles 
shrink into the interior of the alga and merge with one another. 
In practice, the inverse algorithm often works without actually taking 
the alga projection.
One begins with drawing all 1-cycles and tries to shrink merge them 
to obtain a consistent tiling.

\begin{figure}[htb]
\begin{center}
\includegraphics[width=12.0cm]{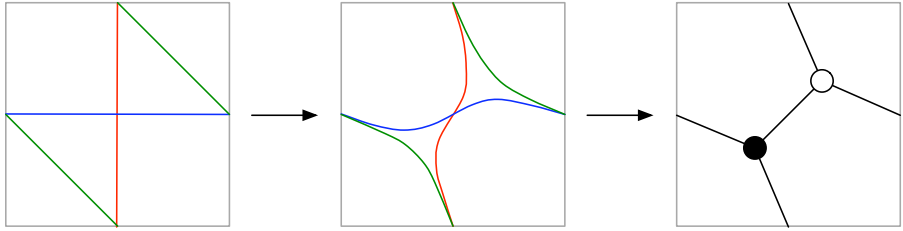}
\caption{Alga and the inverse algorithm for $\IC^3$.} \label{alga}
\end{center}
\end{figure}

\begin{figure}[htb]
\begin{center}
\includegraphics[width=6.0cm]{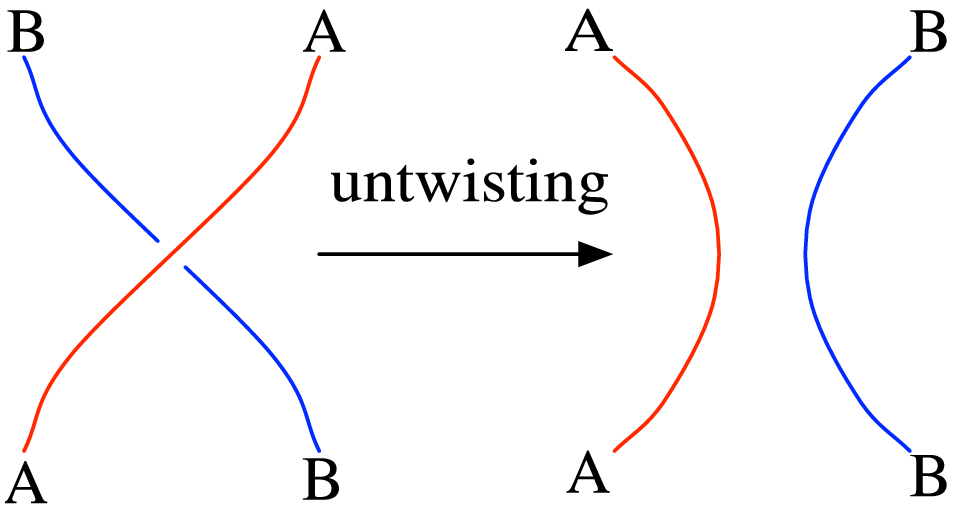}
\includegraphics[width=12.0cm]{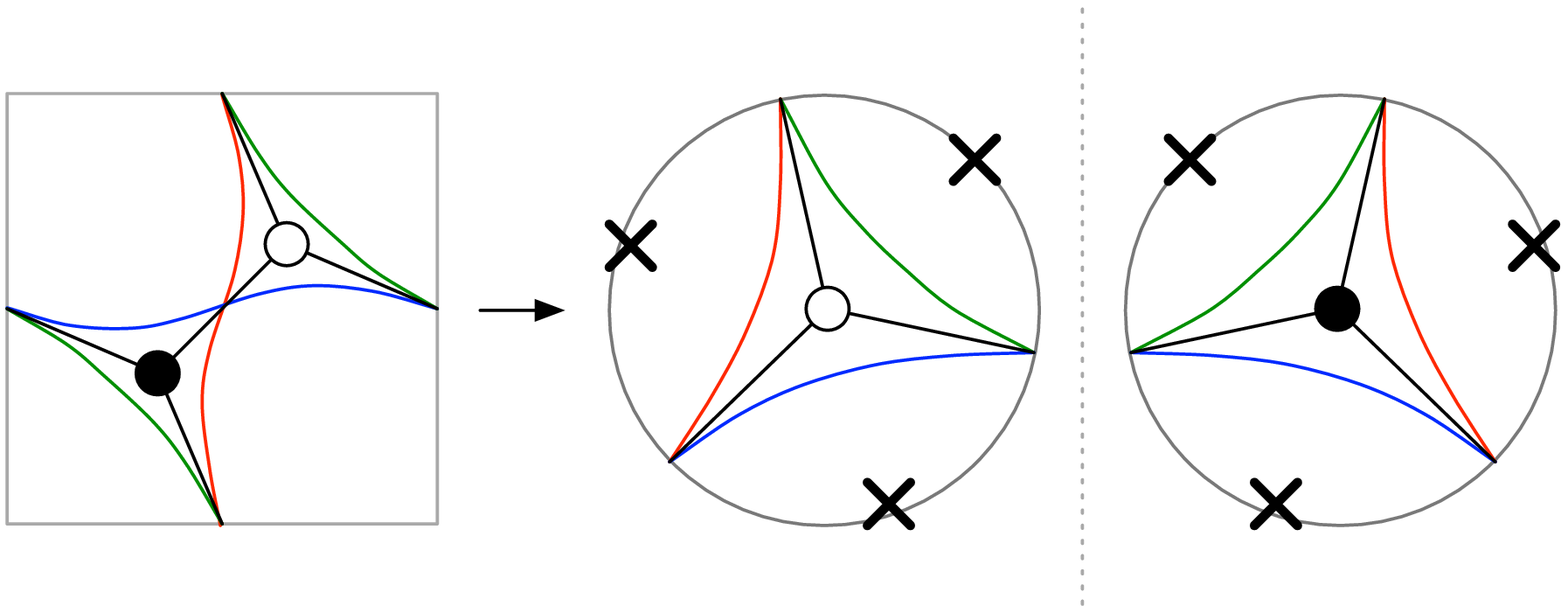}
\caption{Untwisting. Two discs whose boundaries
are identified represent a two-sphere. The crosses denote the
punctures encircled by the 1-cycles.} \label{alam}
\end{center}
\end{figure}

Since the alga and the amoeba are just two different projections 
of the same holomorphic surface, a simple map relating the two 
should exist. The map was found in \cite{t5} and named the untwisting. 
When any two of the 1-cycles intersect in the alga, one locally 
untwists the intersection to obtain a new configuration. 
The result is precisely the amoeba. 
Figure \ref{alam} shows how the untwisting works for $\IC^3$.

\subsection{Geometry of $\S$}

As in the tiling model, we expect that the projections 
of the special Lagrangian manifold $\S$ will play a crucial 
role in the crystal model. As explained in section 2, 
supersymmetry requires that $\S$ be locally 
a product of a plane in $\IR^3$ and a 1-cycle in $T^3$. 
Moreover, in order for $\S$ to be calibrated properly, 
the plane corresponding to a $(p,q,r)$ 1-cycle 
should be orthogonal to a $(p,q,r)$ vector in $\IR^3$.  
Since $\S$ must be completely determined by the toric data, 
it is not difficult to guess what the planes and the 1-cycles are. 
Each edge $w_{IJ} = v_I - v_J$ of the toric diagram 
and the 2-fan orthogonal to the edge naturally give 
the plane/1-cycle pair. 

\begin{figure}[htb]
\begin{center}
\includegraphics[width=4.0cm]{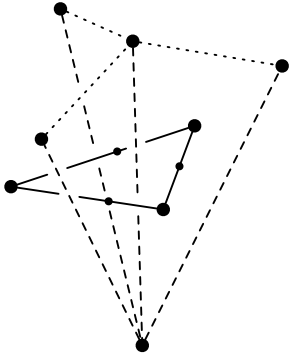}
\caption{A part of the toric diagram. Each edge of 
the toric diagram corresponds to a 2-fan in $\IR^3$ 
and a 1-cycle in $T^3$. Suppressing 
the radial direction, the 2-fans form the dual toric diagram.} \label{2fan}
\end{center}
\end{figure}

Just as the amoeba in the tiling model is a thickened $(p,q)$-web 
and the boundary of the alga gave the $(q,-p)$ 1-cycles, 
we expect that the amoeba and the alga projections 
of the special Lagrangain manifold $\S$ in the crystal model 
will give a thickened 2-fans in $\IR^3$ paired with 1-cycles in $T^3$. 
Unfortunately, to our knowledge, an explicit equation describing the shape of 
the special Lagrangian manifold $\S$ is not known,
\footnote{
We thank D. Joyce for a correspondence on this point.}
but we can get some intuition from the following consideration.

Recall that the holomorphic surface 
for the tiling model was given by the Newton polynomial. 
If we write down the Newton polynomial for a CY$_4$ as in 
(\ref{newton}) but with three $\IC^*$ variables,  
it is locally a product of a plane in $\IR^3$ and a {\em 2-cycle} in $T^3$. 
We now take the {\em mirror} of the $(C^*)^3$, that is, T-dualize along 
the $T^3$-fiber of $(\IC^*)^3$, then we obtain a dual $(\IC^*)^3$. 
If there were a D4-brane wrapping the holomorphic surface in the original 
$(\IC^*)^3$, then it would be transformed into a D3-brane wrapping 
a special Lagrangian submanifold of the dual $(\IC^*)^3$. 
We conjecture that this is precisely the $\S$ we are looking for. 
We should warn the readers that we are using mirror symmetry as a tool 
to relate a manifold to another, and it has nothing to do with 
the physical symmetry of the crystal model. 

In the tiling model, it was useful to represent the amoeba 
as a punctured Riemann surface. The punctures are the 
points at infinity along the legs of the $(p,q)$-web. 
In the crystal model, we can also think of the amoeba of $\S$ 
as a 3-manifold with some defects. As we shrink the 2-fans 
along the radial direction, the `points at infinity' form a locally 
one dimensional defect. The 1-cycles paired with 
the 2-fans are localized along the defects. 
Globally, the defect is isomorphic 
to the dual toric diagram. As we will see in the next subsection,
when the toric diagram has no internal points, 
the amoeba has the topology of a three-sphere apart from the defect. 
It will be important to find out the topology of the amoeba 
in the general case in order to complete the inverse algorithm.

\subsection{The alga and the crystal}

We now put together the contents of the previous two subsections 
to find the inverse algorithm for the crystal model. 
It is natural to expect that the alga projection of $\S$ onto $T^3$ 
will again reveal the structure of the intersection between the $\S$-brane 
and the $T$-branes. With no explicit equation describing $\S$, 
we cannot derive the alga from a first principle. 
Nevertheless, in analogy with the tiling model, we can guess the shape 
of the alga and make consistency checks later.

\begin{figure}[htb]
\begin{center}
\includegraphics[width=12.0cm]{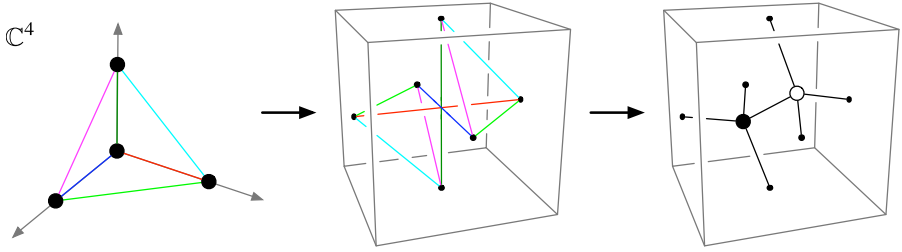}
\caption{Each edge of the toric diagram gives a 1-cycle in the $T^3$. The 1-cycles define certain solids. The 1-cycles bend into the solids and merge to form the bonds and atoms of the crystal.} \label{solid1}
\end{center}
\end{figure}

\begin{figure}[htb]
\begin{center}
\includegraphics[width=12.0cm]{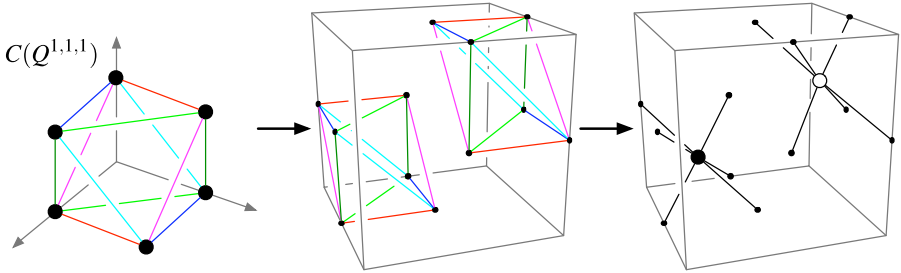}
\caption{The crystal for $C(Q^{1,1,1})$ has the NaCl structure.} \label{solid2}
\end{center}
\end{figure}

Each edge of the toric diagram gives a 1-cycle in the $T^3$. 
Unlike in the tiling model, the 1-cycles generically do not 
intersect with each other. In order to fix the positions of the 1-cycles 
in the $T^3$, we make the following observation.  
When several edges form a boundary of a face of the toric diagram, 
the total homology charge of the corresponding 1-cycles vanish. 
Since the 2-fans meet with each other, the 1-cycles can join each other 
and be deformed smoothly into a vanishing cycle. Therefore, we require that 
the 1-cycles associated to a same face of the toric diagram 
intersect with one another in the $T^3$. In the simplest cases, 
this requirement is sufficient to construct the alga of a given CY$_4$.  
Figures \ref{solid1} and \ref{solid2} show the alga and the crystals for 
$\IC^4$ and $C(Q^{1,1,1})$.

These crystals are particularly simple because  
all lattice points of the toric diagram are corners, that is, 
they do not lie in the interior, on the surfaces, or along the edges.
In fact, when the toric diagram is a convex subset of a unit cube, 
there is a one-to-one correspondence between the vertices 
of the toric diagram and the bonds of the crystal. 
Note that Figure \ref{xamp}(e) also falls into this class.

\begin{figure}[htb]
\begin{center}
\includegraphics[width=12.0cm]{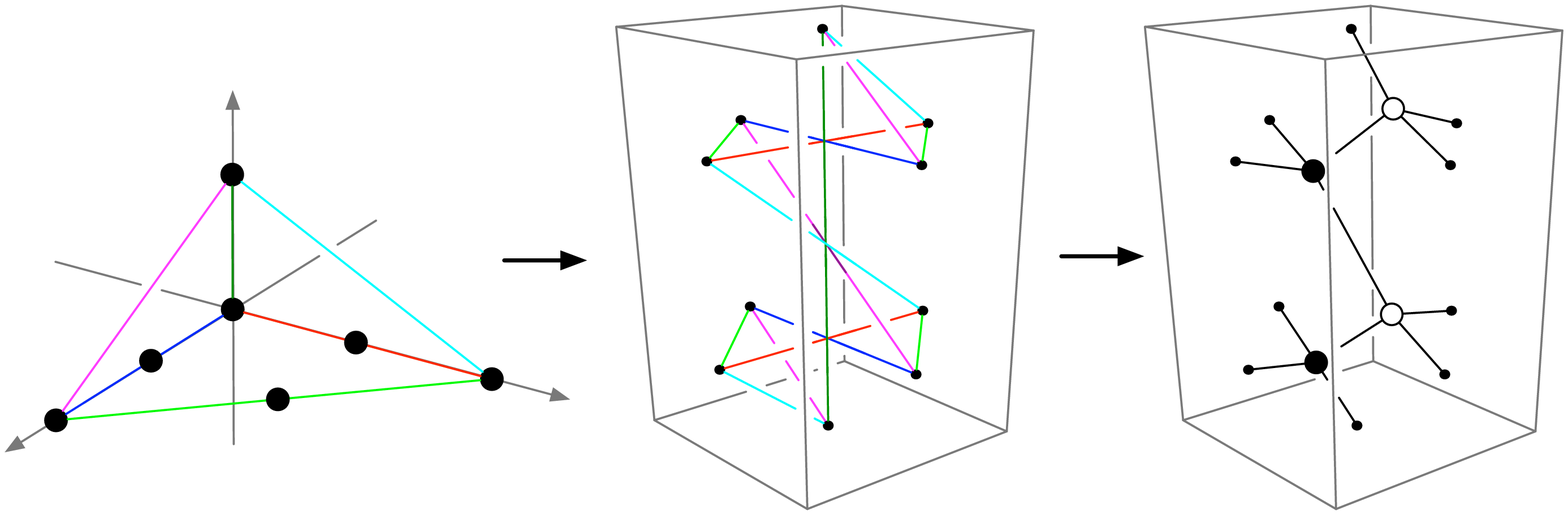}
\caption{The crystal for $(\IC^3/\IZ_2\times \IZ_2)\times \IC$.} \label{solid3}
\end{center}
\end{figure}

Applying the algorithm to orbifolds of $\IC^4$ leads to 
slightly more complicated crystals. Figure \ref{solid3} 
shows the crystal for $(\IC^3/\IZ_2\times \IZ_2)\times \IC$ 
whose toric diagram is in Figure \ref{xamp}(d).
Note that, compared to the $\IC^4$ case, the toric diagram 
is enlarged by a factor of 2 in the $x$ and $y$ directions, 
while the unit cell in the crystal is twice as large in the $z$ direction. 
So, in contrast to the tiling model, it is not straightforward 
to obtain the crystal for $(\IC^3/\IZ_m\times \IZ_n)\times \IC$ for 
arbitrary $m$ and $n$. 

\begin{figure}[htb]
\begin{center}
\includegraphics[width=12.0cm]{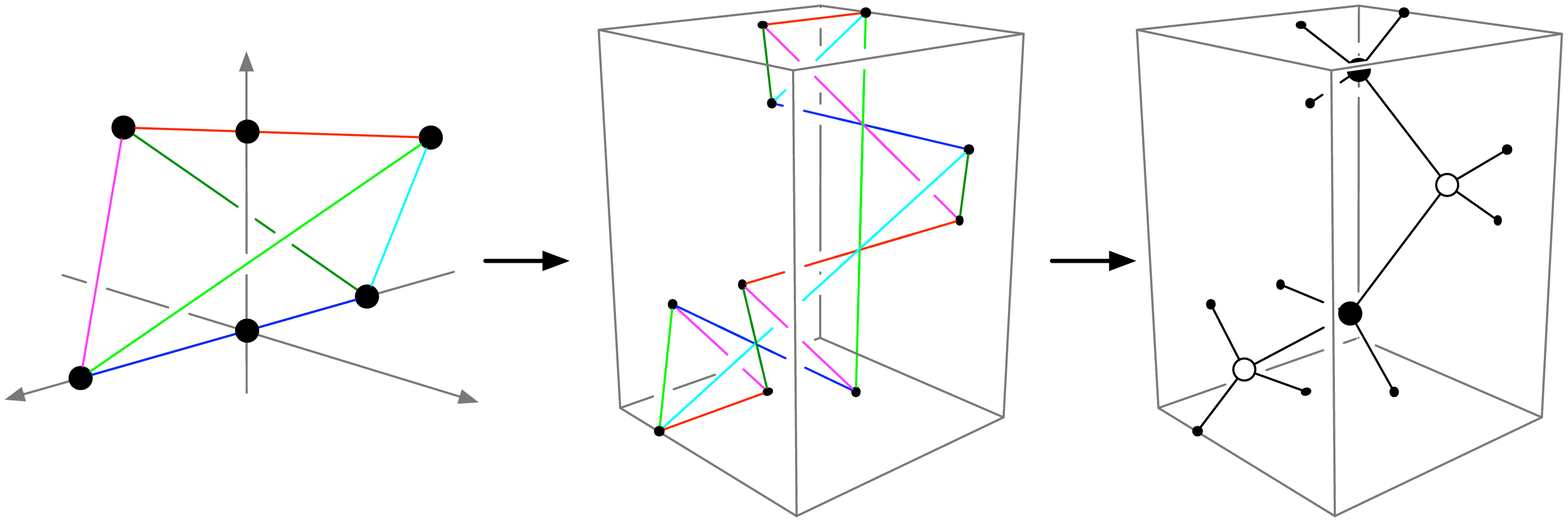}
\caption{The crystal for $(\IC^2/\IZ_2)^2$.} \label{solid4}
\end{center}
\end{figure}

Figure \ref{solid4} shows the crystal for another orbifold,  $(\IC^2/\IZ_2)^2$. 
Again, the crystal structure for the orbifolds is the same as that of $\IC^4$, 
except that the unit cell is twice as large. The basis vectors 
of the new unit cells in terms of those of $\IC^4$ are 
$(0,0,1)$, $(1,-1,0)$, $(1,1,-1)$.

\subsection{Untwisting and the amoeba}

In the tiling model, the untwisitng map reflects the fact that 
the real and imaginary parts of a holomorphic function 
are related to each other through the Cauchy-Riemann condition.
As the special Lagrangian manifold $\S$ is a `mirror' of a holomorphic 
surface, we expect that an analog of the untwisting map 
exists. 

\begin{figure}[htb]
\begin{center}
\includegraphics[width=8.0cm]{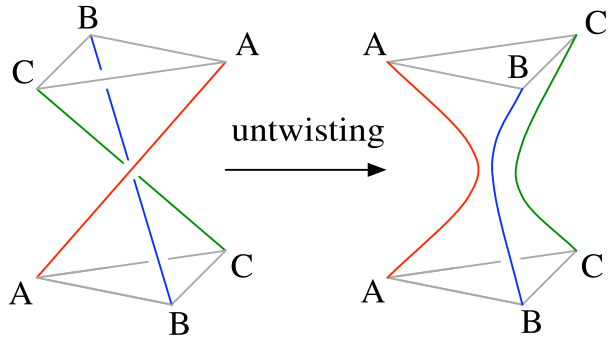}
\caption{Untwisting the crystal.} \label{un3d}
\end{center}
\end{figure}

\begin{figure}[htb]
\begin{center}
\includegraphics[width=13.0cm]{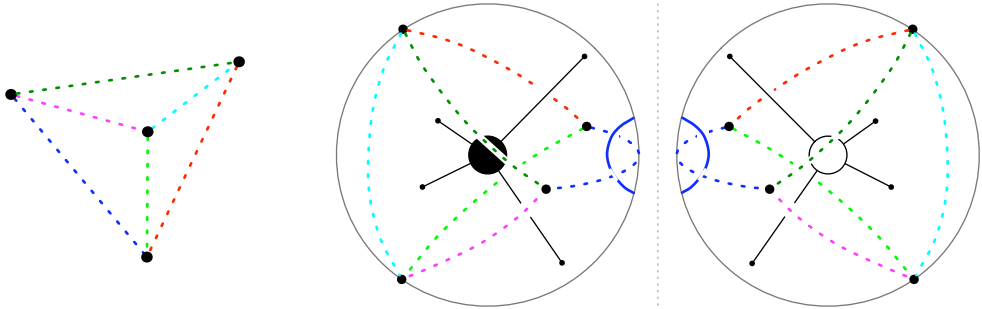}
\caption{Amoeba for $\IC^4$. We represent a three-sphere 
as the union of two balls with the surfaces identified. 
The dotted lines on the balls denote defects.} \label{am2}
\end{center}
\end{figure}

It is indeed possible to untwist the alga 
to obtain the amoeba which has all the properties 
we discussed earlier. Just as in the tiling model, 
the untwisting flips the orientation of the space transverse to 
the bond of the crystal. See Figure \ref{un3d}.
We apply the untwisting map to the alga of $\IC^4$, 
we obtain the amoeba depicted in Figure \ref{am2}. 
Note that the dual toric diagram is a tetrahedron as expected. 
The 1-cycles are localized along 
the dual toric diagram as they should be. Applying the untwisting 
map to $C(Q^{1,1,1})$ yields a similar result, with the dual toric diagram 
being a cube.

The untwisting map for the orbifolds is more non-trivial. 
The dual toric diagrams have double-lines because some points of 
the toric diagrams sit on the edges. It is a priori not clear 
how the double line should split in the amoeba. 
The untwisting map answers the question.

\begin{figure}[htb]
\begin{center}
\includegraphics[width=13.0cm]{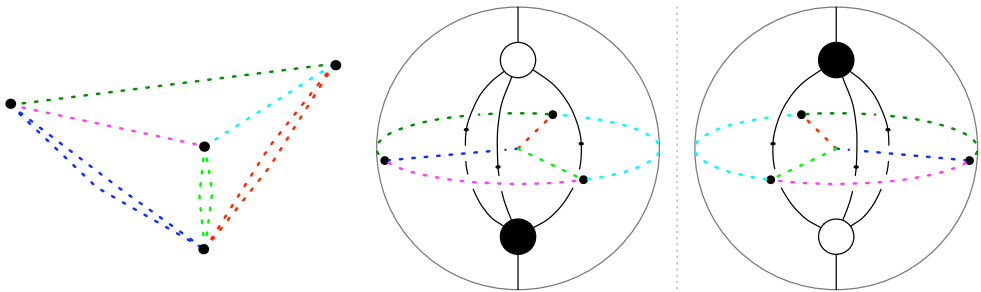}
\caption{The dual toric diagram and its embedding into 
the amoeba for $(\IC^3/\IZ_2\times \IZ_2)\times \IC$.} \label{dual}
\end{center}
\end{figure}

\begin{figure}[htb]
\begin{center}
\includegraphics[width=13.0cm]{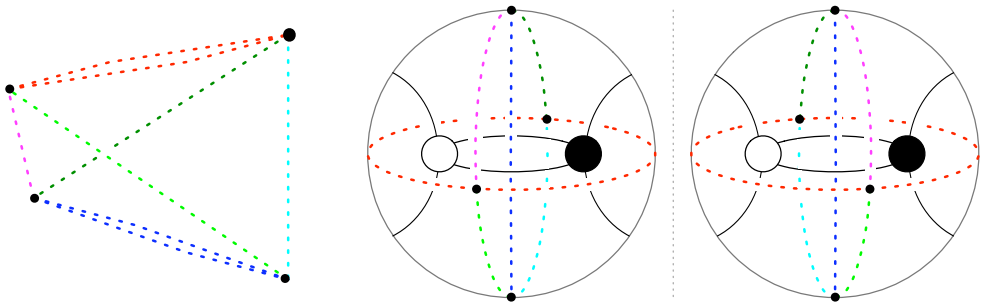}
\caption{The dual toric diagram and its embedding into 
the amoeba for $(\IC^2/\IZ_2)^2$.} \label{am3}
\end{center}
\end{figure}

\section{Forward algorithm}

The forward algorithm of the tiling model as described in \cite{t1} 
is based on two related notions, namely, perfect matchings and 
the Kasteleyn matrix. In this section, we show that 
the same tools work equally well for the crystal model 
with little modification. 
%Since the inverse algorithm for the crystal model is not yet complete, 
%the forward algorithm allows us to guess a crystal for a given CY$_4$ 
%and check whether the guess is correct. This is in fact how we obtained 
%some of the crystals presented in this paper. 

\subsection{Perfect matchings}

The concept of perfect matchings plays a crucial role 
in the forward algorithm. 
A perfect matching is a subset of bonds of the crystal, 
such that every atom of the crystal is an endpoint of precisely one such bond. 
The bonds in each perfect matching carry 
an orientation. We choose to orient the arrows to go from 
a white atom to a black one. 
%The four perfect matchings of the $\IC^4$ crystal is shown in figure XXX. 

In the tiling model, perfect matchings are known to have several nice properties. 
In this section, we show that the same is true of the crystal model.  
We focus on two of the main features:
\bn
\item 
Each perfect matching can be located in the toric diagram. 
The relative coordinate in the toric diagram 
between two perfect matchings $p_\a$ and $p_\b$ 
is given by the homology charge of $(p_\a-p_\b)$ regarded 
as a one-cycle in $T^3$. 
\item 
The perfect matchings solve the `abelian' version of the F-term condition 
for the chiral fields $X_i$ associated to the bonds, if we set 
\be
\label{Fsol}
X_i = \prod_\a p_\a^{\langle X_i, p_\a \rangle},
\ee
where $\langle X_i, p_\a \rangle$ equals 1 if $p_\a$ contains the bond $X_i$ and 0 
otherwise. 
\en
\begin{figure}[htb]
\begin{center}
\includegraphics[width=15.0cm]{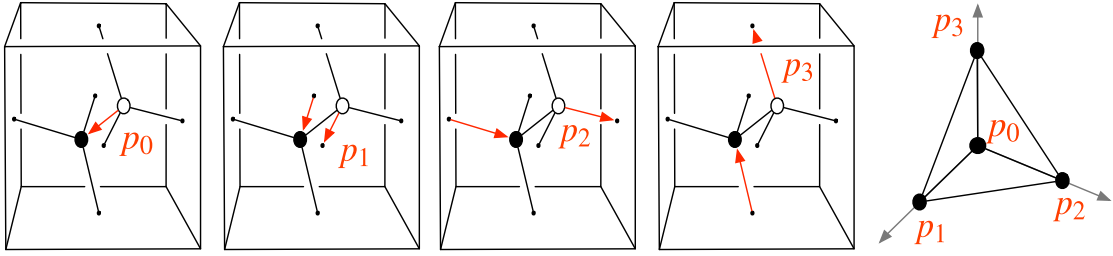}
\caption{Perfect matchings of the crystal for $\IC^4$} \label{C4matching}
\end{center}
\end{figure}

Instead of giving a general proof, we illustrate the idea with simple,
but non-trivial, examples. 
When the crystal contains only two atoms, 
as is the case for $\IC^4$ and $C(Q^{111})$, each and every bond 
in the crystal corresponds to a perfect matching. 
Figure \ref{C4matching} shows the four perfect matchings of the $\IC^4$ crystal 
and how they are associated to points on the toric diagram. 
For any pair of perfect matchings $p_\a$ and $p_\b$, 
the oriented path $(p_\a-p_\b)$ forms a one-cycle in $T^3$. 
The homology charge translates into the relative coordinate 
between the two points corresponding to $p_\a$ and $p_\b$. 
Since there are only two atoms, the F-term condition 
equates the product of all bonds meeting at an atom 
with the same product at the other atom. 
So, the abelian F-term condition is trivially satisfied. 

\begin{figure}[htb]
\begin{center}
\includegraphics[width=15.0cm]{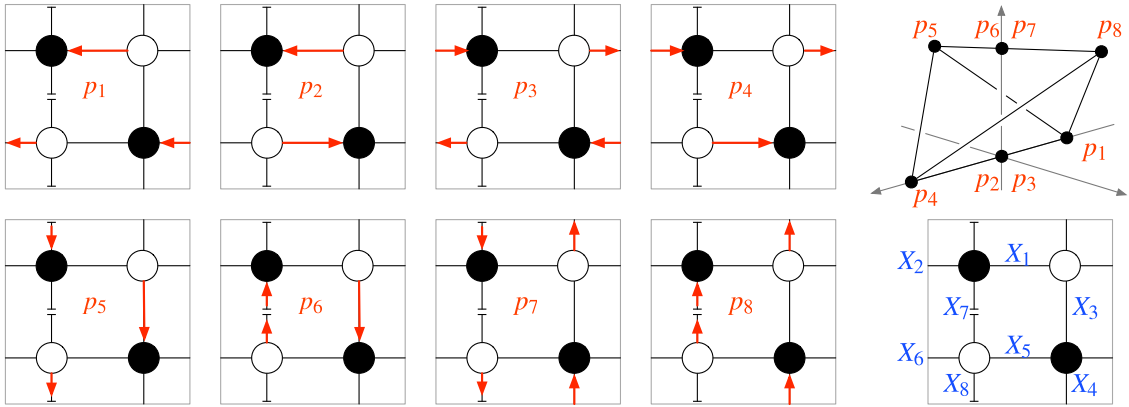}
\caption{Perfect matchings of the crystal for $(\IC^2/\IZ_2)^2$
projected onto the $xy$-plane. The short double lines 
on $X_7$, $X_8$ means that they go over different unit cells 
in the $z$-direction} \label{matching2}
\end{center}
\end{figure}

The map between perfect matchings and points in the toric diagram 
becomes less trivial as the number of atoms increases. 
Figure \ref{matching2} shows the eight perfect matchings of the 
$(\IC^2/\IZ_2)^2$ crystal 
and their location in the toric diagram. 
The F-term condition states that the product of bonds meeting at 
any atom is the same. Labeling the bonds as in the figure, 
the condition reads
\be
X_1 X_2 X_3 X_4 = X_3 X_4 X_5 X_6 = X_5 X_6 X_7 X_8 = X_7 X_8 X_1 X_2 . 
\ee
It is easy to check that (\ref{Fsol}) solves this constraint. Explicitly, 
\be
X_1 = p_1 p_2, \;\; X_2 = p_3 p_4, \;\; X_3 = p_5 p_6, \;\; X_4 = p_7 p_8, \nn \\
X_5 = p_2 p_4, \;\; X_6 = p_1 p_3, \;\; X_7 = p_6 p_8, \;\; X_8 = p_5 p_7.
\ee

The relation (\ref{Fsol}) also allows us to determine the charges 
of the fields $X_i$. The corners of the toric diagram correspond 
to baryons, and their charges can be computed from the volume of 
the associated supersymmetric 5-cycles. If we assign the charges to 
the corresponding perfect matchings and set the charges of other 
perfect matchings to zero, then (\ref{Fsol}) determines the charges 
of the fields $X_i$. Moreover, the super-potential terms contain 
each and every corner precisely once, so that their $R$-charges 
are always 2. 

As a final remark, we note that just as in the tiling model, 
one can use the Kasteleyn determinant as a generating function of 
the perfect matchings. 

\subsection{Relation between the linear sigma model and the associated
  crystal}

From the toric diagram of $C(Q^{1,1,1})$ in Figure 3, one has the
following toric vertices
\be
v_1 =(1,0,0,1), \;\; v_2=(0,1,0,1), \;\; v_3=(1,1,0,1), \nn \\
v_4 =(0,0,1,1), \;\; v_5=(1,0,1,1), \;\; v_6=(0,1,1,1).
\ee
From the relation (\ref{ker}) one can easily find the charges of the
linear sigma model fields $\sigma_I$ corresponding to $v_I$ since in this case 
there are one to one correspondence between the perfect matchings and
the vertices of the toric diagram. The charge vectors for $\sigma_I$
are given by
\begin{eqnarray}
Q_1 &=&(1,1,-1,-1,0,0), \nn \\
Q_2 &=& (0,0,1,1,-1,-1).
\end{eqnarray}
The two charge assignments are the usual ones  for the sigma model
describing $C(Q^{1,1,1})$. 

If we go to more complicated examples such as $\IC^2/\IZ_2 \times
\IC^2/\IZ_2$, we have eight perfect matchings so that the symplectic quotients
are described by $U(1)^4$ gauge theory with suitably charged matters. 
One can obtain the charge matrices for eight perfect matching fields 
again using the relation (\ref{ker}). However there are ambiguities 
in the charge assignments since one can think of various different
$U(1)$ combinations
in describing the same symplectic quotients. In the tiling model, the
charge assignments arise naturally. The linear sigma model fields 
are suitable combinations of bifundamental fields appearing in the 
D-branes located in the considered Calabi-Yau singualrities. 
Here the relation between the fundamental excitations of membrane
and the linear sigma model fields is not clear. We introduce the
fields assigned with bonds and they are related to the prefect
matchings. It is desirable to understand this issue better.

\subsection{Trivalent atoms and non-uniqueness of the crystal}

In the tiling model, a vertex with only two edges correspond to 
a mass term in the gauge theory. Integrating out the mass term 
translates into shrinking the vertex and the two edges away completely 
from the tiling. It is likely that the same is true of the crystal model, 
since as we will discuss in the next section, the atoms of crystals 
are interpreted as super-potential terms. 

Unlike in the tiling model, trivalent atoms of the crystal model 
also deserve special attention. As discussed in section 2, 
in the inverse algorithm, the atoms appear from solid components 
of the alga. A proper three dimensional solid has at least four vertices, 
which means that the resulting atom should have at least 
four bonds. It is still conceivable that trivalent atoms appear 
as the solid is somehow `squashed' to lie in a plane. 
In fact, we have found several crystals with trivalent atoms. 
Although the inverse algorithm suggests that they are rather unnatural, 
so far we have not found any physical inconsistency to rule them out. 

\begin{figure}[htb]
\begin{center}
\includegraphics[width=4.5cm]{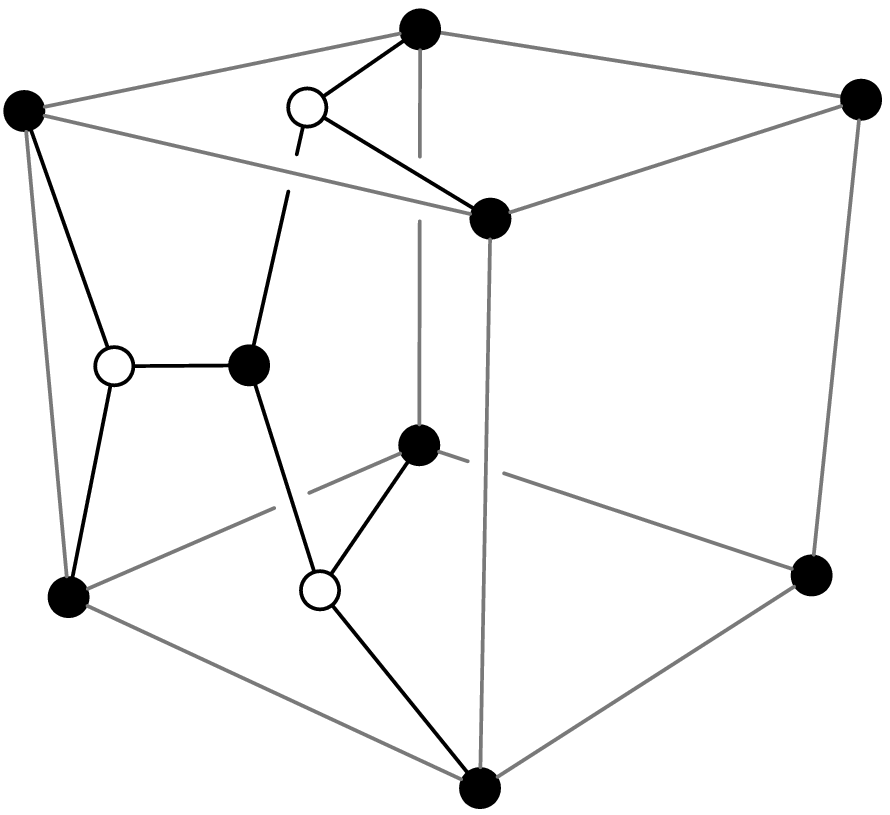}
\caption{The crystal for $\IC^2/Z_2\times \IC^2$.} \label{Z2}
\end{center}
\end{figure}

The first example of a crystal with trivalent bonds comes from 
the orbifold $\IC^2/\IZ_2 \times \IC^2$ whose toric diagram is given 
in Figure \ref{xamp}(b) and crystal depicted in Figure \ref{Z2}.
It is easy to check that the crystal can reproduce the toric diagram with the
perfect matching. The R-charges can be assigned to the bonds as before.
%When we project our crystal on the $12$-plane,
%those trivalent atoms seem to be mass terms in the dimer model
%\footnote{We are in fact not sure that those projections give
%proper information.}. 
%We however cannot shrink those atoms simply
%as the dimer model because of the mismatch between white and black atoms. 

\begin{figure}[htb]
\begin{center}
\includegraphics[width=8.5cm]{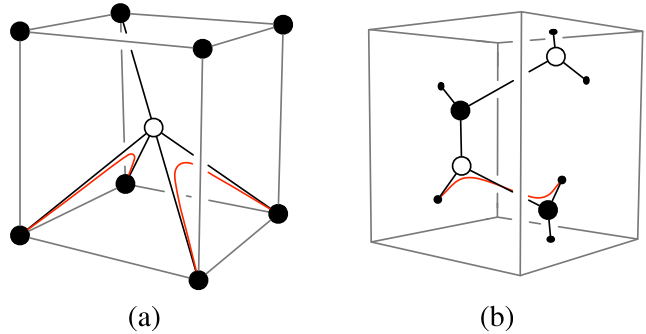}
\caption{The two crystals for $C(T^{1,1})\times \IC$.} \label{T11}
\end{center}
\end{figure}

Another interesting example is $C(T^{1,1}) \times \IC$. 
Its toric diagram is described in Figure \ref{xamp}(e).
This is our first example having two different crystals; see Figure \ref{T11}.  
The crystal (a) is derived from the inverse algorithm just 
as the $\IC^4$ and $C(Q^{1,1,1})$ examples discussed in section 3, 
and does not contain any trivalent atom. 
The crystal (b), on the other hand, contains trivalent atoms only. 
Both crystals give the same toric diagram as far as the perfect 
matching method is concerned. 

To determine whether both crystals are physically relevant 
will require a more thorough understanding of the inverse algorithm.
In the tiling model, two or more tilings corresponding to 
the same toric diagram are known to be related to each other by 
Seiberg duality. So, if two or more crystals are indeed allowed, 
we will have to find out an analog of Seiberg duality in the new setup. 

The crucial difference between two crystals is that the crystal (a) 
has two homology $(1,0,0)$ 1-cycles as shown in the Figure \ref{T11} 
(red curves), but the other one has only one such 1-cycle. 
It can be understood that two 1-cycles in the crystal (a) are degenerate in 
the crystal (b). 
In the tiling model, degenerate 1-cycles often lead to inconsistent dimer graphs. 
Even though degenerate cycles do not lead to an inconsistent crystal, 
we may say that the crystal containing trivalent
atoms is more singular based on the degenerate 1-cycles.  

%\subsection{A proposal for $M^{32}$}

\begin{figure}[htb]
\begin{center}
\includegraphics[width=4.5cm]{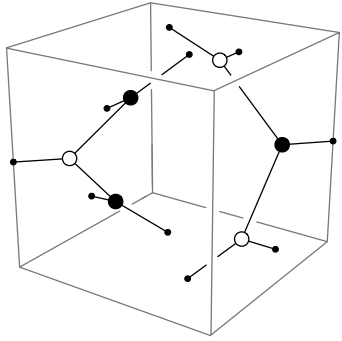}
\caption{A candidate for the $C(M^{3,2})$ crystal.} \label{M32}
\end{center}
\end{figure}

\begin{figure}[htb]
\begin{center}
\includegraphics[width=13.0cm]{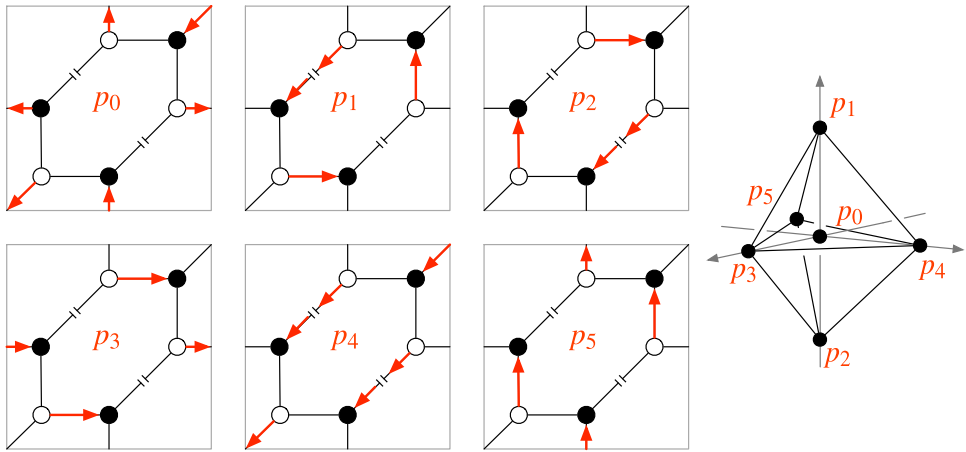}
\caption{Perfect matchings of the $C(M^{3,2})$ crystal candidate, 
projected onto the $xy$-plane.} \label{M32m}
\end{center}
\end{figure}

We close this subsection with a proposal for the crystal structure 
of $C(M^{32})$, whose toric diagram
is given in Figure \ref{xamp}(g). 
With our limited understanding of the inverse algorithm, we have 
not been able to derive the crystal for $C(M^{3,2})$ from the toric diagram. 
It is still possible to do some guesswork to find a candidate 
and check whether it gives the correct toric diagram using the perfect matchings. 
Figure \ref{M32} shows one such candidate. As shown in Figure \ref{M32m}, 
the perfect matchings do yield the toric diagram of $C(M^{3,2})$.

However, the crystal consists of trivalent atoms only. 
If we check the $R$-charges of the bonds of the crystal, 
all six edges of the `hexagon' in the interior of the unit cell 
are contained in degenerate 1-cycles in the sense we discussed above. 
There may exist a more regular crystal without trivalent atoms. 
We hope to resolve this issue in a future work.

\section{Counting BPS meson states}

Counting BPS states is one of the most basic problems in AdS/CFT with 
supersymmetry. Very much progress has been made recently in counting 
states preserving a half of the supersymmetry in $D=4$, $\CN=1$ 
\cite{minwalla1, man1, msy2, pleth1, ms2, man2, baryon1, 
romel, grant, baryon2, pleth2} 
or $D=3$, $\CN=2$ theories \cite{man2, minwalla2}. 
Exact partition functions for mesons and baryons have been written down. 
In the crystal model, the BPS meson states are represented by  
closed membrane configurations. 
In this section, we explain in detail how the crystal model 
correctly reproduces the meson spectrum computed 
on the geometry side.

\subsection{Character functions} 

Chiral mesons in CFT$_3$ correspond to algebraic (holomorphic polynomial) functions on the CY$_4$ cone $X$. They are labeled by integer 
points $m=(m_1,m_2,m_3,m_4)$ in the solid cone $\D$, 
which are the momentum quantum numbers 
along the $T^4$ fiber of $X$.
It follows immediately that the flavor charges of a meson $m$ are $F_i(m) = m_i$. 
The $R$-charge of the meson is then given by \cite{zaff, msy2, gmsy}
\be
R(m) 
= \frac{1}{2} b^i F_i(m) = \frac{1}{2} (m \cdot b).
\ee
The spectrum of all mesons is conveniently summarized 
in the character function \cite{msy2, pleth1}. 
\be
Z(q_i;X) \equiv \sum_{\{m \}} \prod_{i=1}^{4} q_i^{m_i}
\ee
When the CY$_4$ is toric, a closed form of the character function 
can be read off from the toric diagram. 
One draws the graph dual of a triangulation of the toric diagram. 
The result is a tetra-valent graph whose external legs are the 1-fans 
discussed in section 2. Then the partition functions is given by
\be
Z(q_i;X) = \sum_{\a \in T} \prod_{b=1}^{4} \frac{1}{1- \prod q_i^{a^i_{\a b}}}.
\ee
The sum runs over the vertices of the graph or, equivalently, the 
simplexes of the triangulation. The product runs over the four edges 
$\vec{a}_{\a b}$ meeting at the vertex $\a$. See \cite{msy2, pleth1} 
for more details.
Applying this formula to the examples of the current paper, we find
\be
\IC^4 &:& 
\frac{1}{(1- q_1 )(1-q_2)(1- q_3)(1- q_4/q_1 q_2 q_3)},
\\
\IC^2/\IZ_2 \times \IC^2  &:& 
\frac{ 1+ q_4/q_1 q_2}{(1-q_1)(1-q_2)( 1 - q_3 q_4/q_1 q_2)(1- q_4/q_1 q_2 q_3 )},
\\
(\IC^2/\IZ_2)^2  &:&  
\frac{(1+ q_3) (1 + q_4/q_3) }{(1-q_1q_3)(1-q_3/q_1)(1-q_2q_4/q_3)(1-q_4/q_2q_3)},
\\
\IC^2/(\IZ_2 \times \IZ_2) \times \IC  &:&  
\frac{1 + q_4/q_3}{ (1-q_1)(1- q_2)(1- q_3)(1- {q_4}^2 / q_1 q_2 {q_3}^2  )},  
\\
C(T^{1,1}) \times \IC  &:& 
\frac{1-q_4/q_3}{(1-q_1)(1-q_2)(1-q_3)(1-q_4/q_1 q_3)(1-q_4/q_2 q_3)},
\ee
and similar expressions for $C(Q^{1,1,1})$ and $C(M^{3,2})$.
%\be
%\frac{1}{(1 - q_1)(1 - \frac{q_4}{q_2})(1 - \frac{q_2}{q_3})( 1- \frac{q_3}{q_1})} 
%&+& \frac{1}{(1- \frac{q_4}{q_1})(1 - q_2)(1-\frac{q_1}{q_3})(1-\frac{q_3}{q_2})} 
%\nn \\
%+ \frac{1}{(1- \frac{q_4}{q_3})(1- \frac{q_1 q_2}{q_3})(1- \frac{q_3}{q_1})(1- \frac{q_3}{q_2})} 
%&+& \frac{1}{(1 -\frac{q_1}{q_3})(1- \frac{q_2}{q_3})(1 - q_3)(1 - \frac{q_3 q_4}{q_1 q_2})}
%\nn
%\ee
%\\ \hline
%$\mathcal{C}(\mathrm{M}^{3,2})$ 
%& $\begin{matrix}\frac{1}{(1 - q_1)(1 - \frac{1}{q_2})(1 - \frac{q_2 q_4}{q_1 q_3})( 1- q_3)} + \frac{1}{(1- q_2)(1 - q_1 q_2)(1-\frac{q_4}{q_1 {q_2}^2 q_3})(1- q_3)}  \\ + \frac{1}{(1 - \frac{1}{q_1})(1 - \frac{1}{q_1 q_2})(1-\frac{{q_1}^2 q_2 q_4}{q_3})(1-q_3)} + \frac{1}{(1- q_1)(1- \frac{1}{q_2})(1- \frac{1}{q_3})(1- \frac{q_2 q_3 q_4}{q_1})} \\ + \frac{1}{(1 - q_2)(1 - q_1 q_2)(1 - \frac{1}{q_3})(1 - \frac{q_3 q_4}{q_1 {q_2}^2})} + \frac{1}{(1 - \frac{1}{q_1})(1 -\frac{1}{q_1 q_2})(1 - \frac{1}{q_3})(1 - {q_1}^2 q_2 q_3 q_4)}\end{matrix}$ 
The volume of the base of the cone, $Y$, 
can be obtained from the character functions \cite{msy2}.
\be
\frac{\mbox{Vol}(Y)}{\mbox{Vol}(S^7)} 
=  \lim_{t\goto 0} t^4 Z(e^{-b^i t};X=C(Y)).
\ee
The value of the Reeb vector which renders $Y$ Einstein 
(or, equivalently, $X$ Ricci-flat), can be determined by 
minimizing the volume \cite{msy1}. 
Inserting the value back to the character function by $q_i\goto
q^{\frac{b^i}{2}}$, 
we find the level surface, that is, we can count the degeneracy of 
mesons at each value of the $R$-charge. For $\IC^4$ and its orbifolds, 
the result is: 
\be
 \IC^4 : 
 \frac{1}{(1-q^{1/2})^4} ,
%=1+ 4q^{\frac12} + 10q + \cdots  
&\qquad&
 \IC^2/\IZ_2 \times \IC^2 :  
\frac{1+q}{(1-q)^2 (1-q^{1/2})^2} ,
%=1+ 2q^{\frac12} + 6q+ 10q^{\frac32} + 19q^2 + \cdots  
\\
(\IC^2/\IZ_2)^2 : 
\frac{(1+q)^2}{(1-q)^4},
%= 1+ 6q + 19q^2+ 44q^3 + 85q^4 + \cdots  
&\qquad&
(\IC^3/\IZ_2 \times \IZ_2) \times \IC :
\frac{1+q^{3/2}}{(1-q)^3(1-q^{1/2})}.
%= 1+ q^{\frac12} + 4q+ 5q^{\frac32} + 11q^2 + \cdots  
\ee
For other examples we have been studying, the answer is:
\be
C(T^{1,1}) \times \IC  
&:&  \frac{1+q^{3/4}}{(1-q^{3/4})^3(1-q^{1/2})}
=1+ q^{\frac12} + 4q^{\frac34}+ \cdots , 
\\
C(Q^{1,1,1})  
&:&  \frac{1+4q+q^2}{(1-q)^4}= 
\sum_{k} (k+1)^3 q^k ,
%= 1+ 8q + 27q^2 + \cdots  
\\
C(M^{3,2})  
&:&  \frac{(1+q^2)(1+25q^2+q^4)}{(1-q^2)^4}
= \sum_{k} (2k+1)\frac{(3k+1)(3k+2)}{2} q^{2k} .
%= 1+ 30q^2 + 140q^4  + \cdots 
\ee
The last two results agree with other methods used previously to obtain the spectrum \cite{fab}. 

\subsection{Mesons as closed membranes}

Consider an atom with several bonds. 
If we add up all the M2 discs localized along the bonds 
with a fixed orientation, we obtain a spherical membrane 
surrounding the atom. 
In analogy with 
the tiling model, we identify such a configuration as 
a super-potential term of the CFT$_3$.  
The $R$-charge and the flavor charges of the spherical M2-brane 
is the sum of the charges of the component M2-discs. 
The super-potential terms should have $R$-charge two and vanishing flavor charges. 
As pointed out in \cite{crystal}, it is true because (i) every atom corresponds 
to a partition covering the entire toric diagram (ii) the sum of charges 
over all 3-fans is either two ($R$-charge) or zero (flavor charges) 
due to the toric relations.

\begin{figure}[htb]
\begin{center}
\includegraphics[width=13cm]{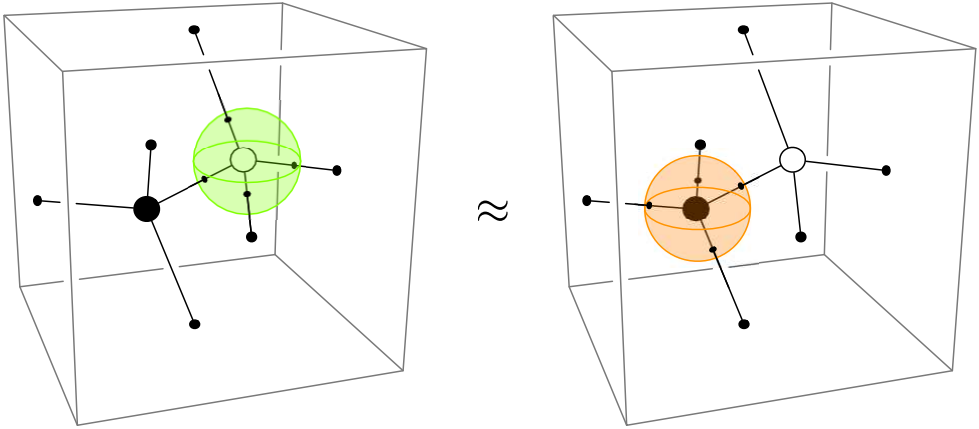}
\caption{Super-potential terms as spherical M2-branes 
surrounding the atoms. The F-term condition 
equates a closed M2-brane wrapping a white atom 
to another M2-brane wrapping a black atom with the opposite orientation.} 
\label{s-potential}
\end{center}
\end{figure}

The super-potential terms yield the F-term equivalence relations among the 
elements of the chiral ring. 
As in the tiling model, the relations imply that the 
sum of terms corresponding to any two connected atoms 
are F-term equivalent to zero. 
%(Why???){\it I guess this is just an
%  assumption which is justified by the correct BPS meson countings and relation to
%the linear sigma model}.
Put another way, a spherical M2-brane wrapping a white atom 
in the crystal is F-term equivalent to another M2-brane wrapping 
a black atom with opposite orientation as illustrated in Figure \ref{s-potential}.
This is why we painted the atoms in two colors in the first place. 
Hence the tiling model and the crystal model share the notions 
of a bipartite graph and dimers.

We make a short digression to what the crystal model implies 
for the field theory description of CFT$_3$. 
An attempt was made in \cite{fab} to write down quiver gauge theories 
dual to AdS$_4\times Q^{1,1,1}$. The proposed theory contained six 
chiral superfields $(A_1,A_2, B_1,B_2, C_1,C_2)$, where $A_i$, $B_i$, $C_i$ 
are doublets under the three $SU(2)$ global symmetries. 
It was shown that the Kaluza-Klein spectrum and the baryon spectrum 
can be reproduced from these fields if one assumes a certain symmetrization rule. 
In CFT$_4$, the symmetrization rule follows from the super-potential 
through the F-term condition. The authors of \cite{fab} found that 
the unique candidate for the super-potential with the correct quantum numbers 
vanishes identically. 
The crystal model reveals the origin of the problem. 
The crystal for $C(Q^{1,1,1})$ has precisely six fundamental M2-discs 
that matches the field content mentioned above. 
Now, the M2-discs form the super-potential terms over a sphere. 
This geometry cannot be represented by a trace of product of matrices; 
matrices represent strings with two end-points.
The correct description of CFT$_3$ would require an algebraic expression 
beyond matrices. 

Recall that the mesons of CFT$_3$ are KK momentum modes 
in AdS$_4\times Y$, and are labeled by integer points 
$m$ in the solid cone $\D$.
T-duality transforms the mesons into closed M2-branes. 
As shown in \cite{crystal},
the first three components of $m$ define the homology charge 
of the 2-cycle in the $T^3$ that the M2-meson is wrapping. 
%It is instructive to confirm this fact 
%in the crystal model.  
%When an M2-meson wraps a 2-cycle, its charge can be computed 
%from its intersections with the bonds in the crystal, 
%as the meson is a bound state of the fundamental M2-discs. 
%Since the bonds are made of $(p,q,r)$-cycles corresponding to 
%the edges $w_{IJ}$ of the toric diagram, we may equally well compute the  
%charges from the intersection of the meson with the $(p,q,r)$-cycles.
%Following Ref. \cite{buttinew}, we use the fact that $\sum_I F_i^I = 0$ 
%($i \ne 4$) to assign variables $F_i^{IJ}=-F_i^{JI}$ to the edges $w_{IJ}$, 
%such that 
%$F_i^I = \sum_{J} F_i^{IJ}$, where the sum runs over the neighboring vertices. 
%Then we find that the charge of the M2-meson is indeed given by 
%\be
%F_i (m) &=& \sum_{(IJ)} (m\cdot w_{IJ}) F_i^{IJ} 
%= \sum_{(IJ)} m \cdot(v_I-v_J) F_i^{IJ} 
%\nn \\ 
%&=& \sum_I (m \cdot v_I) F^I_i  = m_i . 
%\ee
The last component of $m$, contributing $2m_4$ to the $R$-charge, 
measures how many times the M2-meson wraps a super-potential term. 
The F-term condition ensures that all the M2-meson with the same value of $m$ 
are F-term equivalent, regardless of the precise way they wrap the atoms. 
So there is a unique meson for each value of $m$, 
in accord with the results from the geometric side we reviewed in 
the last subsection. 

\begin{figure}[htb]
\begin{center}
\includegraphics[width=15cm]{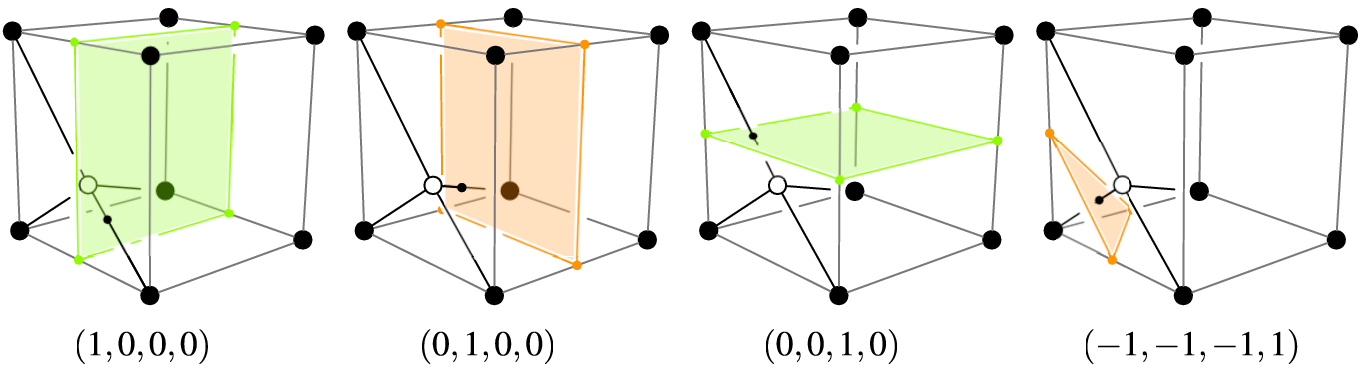}
\caption{Mesons of $\IC^4$ with $R=1/2$.} \label{C4meson}
\end{center}
\end{figure}

It is instructive to check the main assertions of this subsection 
in explicit examples. 
The mesons of $\IC^4$ at $R=k/2$ form a $k$-fold totally symmetric 
tensor representation of the $SU(4)$ global symmetry. 
At $k=1$, the four mesons 
are $m=(1,0,0,0)$, $(0,1,0,0)$, $(0,0,1,0)$ and $(-1,-1,-1,1)$. 
Figure \ref{C4meson} depicts the 2-cycles representing the mesons, 
where we again use colors to denote the orientation of the 2-cycles. 

\begin{figure}[htb]
\begin{center}
\includegraphics[width=15cm]{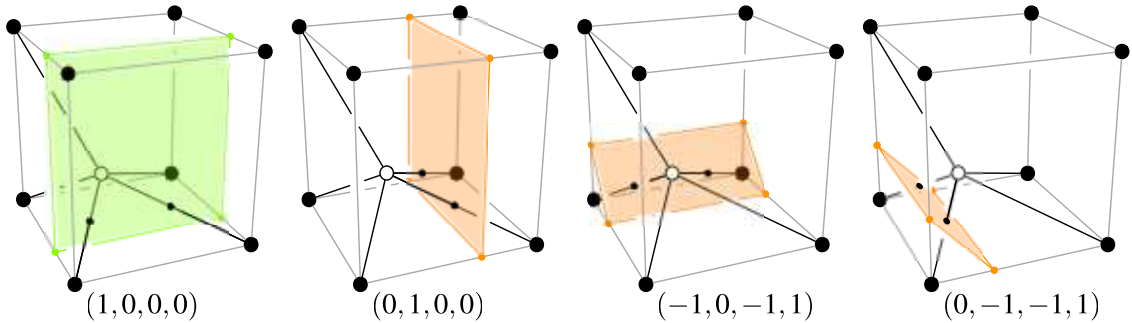}
\caption{Mesons of $C(T^{1,1})\times \IC$ with $R=3/4$.} \label{T11meson}
\end{center}
\end{figure}

For $C(T^{1,1})\times \IC$, the lowest lying meson has $m=(0,0,1,0)$ 
and $R=1/2$. 
It has the same shape as the third meson in Figure \ref{C4meson}. 
At the next level, we have four states with $R=3/4$. 
They are shown in Figure \ref{T11meson}.

So far, we have considered only the single meson states 
with arbitrary values of $m$. In gauge theory terms, 
we have concentrated on `single trace' meson operators in the large $N$ limit. 
It would be interesting to include baryons and take the finite $N$ effect 
into account. 

\section{Discussion}

In this paper, we studied in some detail 
the map between a toric CY$_4$ and the corresponding crystal model. 
Compared to the tiling model, the map is incomplete in many aspects. 
One of the most important gap to fill in is an explicit description 
of the special Lagrangian manifold $\S$. If found, it will prove (or disprove) rigorously the assertions concerning the inverse algorithm and possibly 
reveal more structures. It may also answer some of the questions 
we encountered such as the role of trivalent atoms and 
the relation between different crystals for the same CY$_4$.

It is by now well-known that a few simple formulas hold for the tiling model.
The number of tiles is twice the area of the toric diagram.  
The genus of the amoeba is equal to the number of internal points of the toric diagram. 
The total number of edges of the tiling is given by 
\be
\sum_{I<J} \left| \det\begin{pmatrix} p_I & q_I \cr p_J & q_J \end{pmatrix} \right| , 
\ee
where the sum runs over all legs of the $(p,q)$-web. 
It will be nice to find out whether these relations have 
a counterpart in the crystal model. 

We considered only a limited number of examples in this paper. 
It is certainly desirable to study more examples and relations among them. 
In the study of the tiling model, partial resolutions of a CY$_3$ to obtain 
another less singular CY$_3$ has proved quite useful. 
In the simplest case, partial resolution in the tiling model 
removes some of the edges of the tiling. 
A preliminary study shows that the same is true of the crystal model, 
but a systematic method is yet to be developed. 
As in \cite{uranga}, it is likely that the amoeba projection and the untwisting map 
will again play an essential role. 

Marginal deformation of the CFT$_3$ is another important subject. 
In \cite{lunin}, the derivation of the $\b$-deformed geometry of toric CY$_4$ 
cones tacitly anticipated the crystal model for CFT$_3$. The same paper 
also explained how to interpret the phase factors attached to each 
terms in the super-potential using a sort of $\star$-product. 
The $\star$-product involves the flavor charges of the bi-fundamental fields. 
It would be interesting to give a similar interpretation 
to the $\b$-deformation of CFT$_3$ \cite{lunin, ahn2, glmw}. 
To do so, we will need to define an analog of the $\star$-product 
for membranes, perhaps along the line of \cite{ho}.

\acknowledgments

It is our pleasure to thank Yosuke Imamura, Ken Intriligator, Seok Kim, 
Gautam Mandal, Yutaka Matsuo, Shiraz Minwalla, Soo-Jong Rey, Yuji Sugawara, 
Masahito Yamazaki and Ho-Ung Yee for stimulating discussions, 
and Dominic Joyce for a correspondence. 
Sangmin Lee is grateful to the string theory groups at University of Tokyo and 
Tata Institute of Fundamental Research for hospitality during his visits. 
The work of Sagmin Lee was supported by the Research Settlement Fund 
for the new faculty of SNU and the KOSEF Basic Research Program, 
grant No. R01-2006-000-10965-0. 
Jaemo Park appreciates hospitality of the theoretical particle physics group of
University of Pennsylvania during his stay. 
The work of Jaemo Park is supported by
the Korea Science and Engineering Foundation(KOSEF) grant
R01-2004-000-10526-0
and by the Science Research Center Program of KOSEF through the Center
for Quantum Spacetime(CQUeST) of Sogang University with the grant
number R11-2005-021.
% K. Narayan, S. Minwalla, anyone else??? 

\end{document}